\documentclass[conference,10pt]{IEEEtran}

\usepackage[utf8]{inputenc}
\usepackage[T1]{fontenc}
\usepackage[english]{babel}
\usepackage{cite}
\usepackage{mathtools,amsmath,amssymb}
\usepackage{tikz,pgfplots}
\usepackage{fancyhdr,textcase}
\usepackage{lastpage}
\usepackage{booktabs, cellspace}
\usepackage{acronym}
\usepackage{multirow}
\usepackage{hhline}
\usepackage{csquotes}
\usepackage{url}
\usepackage{setspace}

\usepackage{textcomp}
\usepackage{tasks}
\usepackage{float}
\usepackage{enumitem}
\usepackage{multicol}
\usepackage{listings}
\usepackage{subcaption}
\usepackage{xcolor}
\usepackage[normalem]{ulem}
\usepackage{graphicx}
\usepackage{multicol}

\usepackage[hidelinks]{hyperref}
\usepackage{scalerel}

\usepackage{shellesc} 

\usepackage{tikz,pgfplots}
\pgfplotsset{compat=newest}
\pgfdeclarelayer{background}
\pgfsetlayers{background,main}
\usetikzlibrary{arrows,calc,shadows,positioning,shapes.geometric,decorations.pathmorphing,decorations.shapes,decorations.markings,patterns,svg.path}
\usepgfplotslibrary{groupplots}

\usetikzlibrary{external} 
\newcommand{\secref}[1]{\mbox{Section~\ref{#1}}}
\newcommand{\figref}[1]{\mbox{\figurename~\ref{#1}}}
\newcommand{\tabref}[1]{\mbox{\tablename~\ref{#1}}}

\definecolor{MNO3color}{named}{red}%
\definecolor{MNO2color}{rgb}{0,0.53333,0.02745}
\definecolor{MNO1color}{named}{blue}%
\definecolor{COTScolor}{rgb}{0.85000,0.32500,0.09800}
\definecolor{MECcolor}{rgb}{0.45490,0.26666,0.13333}
\definecolor{mycolor6}{rgb}{0,0.80784314,0.81960784}%

\definecolor{orcidlogocol}{HTML}{A6CE39}
\tikzset{
  orcidlogo/.pic={
    \fill[orcidlogocol] svg{M256,128c0,70.7-57.3,128-128,128C57.3,256,0,198.7,0,128C0,57.3,57.3,0,128,0C198.7,0,256,57.3,256,128z};
    \fill[white] svg{M86.3,186.2H70.9V79.1h15.4v48.4V186.2z}
                 svg{M108.9,79.1h41.6c39.6,0,57,28.3,57,53.6c0,27.5-21.5,53.6-56.8,53.6h-41.8V79.1z M124.3,172.4h24.5c34.9,0,42.9-26.5,42.9-39.7c0-21.5-13.7-39.7-43.7-39.7h-23.7V172.4z}
                 svg{M88.7,56.8c0,5.5-4.5,10.1-10.1,10.1c-5.6,0-10.1-4.6-10.1-10.1c0-5.6,4.5-10.1,10.1-10.1C84.2,46.7,88.7,51.3,88.7,56.8z};
  }
}

\newcommand\orcidicon[1]{\href{https://orcid.org/#1}{\mbox{\scalerel*{
\begin{tikzpicture}[yscale=-1,transform shape]
\pic{orcidlogo};
\end{tikzpicture}
}{|}}}}

\begin{document}

\title{QoS Evaluation and Prediction for C-V2X Communication in Commercially-Deployed \\ LTE and Mobile Edge Networks}

\author{\IEEEauthorblockN{Luis Torres-Figueroa, Henning F. Schepker and Josef Jiru}
\IEEEauthorblockA{Fraunhofer Institute for Cognitive Systems IKS, Munich, Germany}
{\{luis.torres.figueroa, henning.schepker, josef.jiru\}}@iks.fraunhofer.de\\[1mm]}
\maketitle

\begin{abstract}


Cellular vehicle-to-everything (C-V2X) communication is a key enabler for future cooperative automated driving and safety-related applications. The requirements they demand in terms of Quality of Service (QoS) performance vary according to the use case. For instance, Day-1 applications such as Emergency Brake Light warning have less strict requirements than remote driving. In this paper, we seek to answer two questions: Are current LTE networks ready to support Day-1 applications at all times? And, can underperforming situations be reliably predicted based on GPS and network-related information? To address these questions, we first implement a system that collects positioning data and LTE key performance indicators (KPIs) with a higher time resolution than commercial off-the-shelf LTE modems, while simultaneously measuring the end-to-end (E2E) delay of an LTE network. We then use this system to assess the readiness of multiple mobile network operators (MNOs) and a live Mobile Edge Computing (MEC) deployment in an urban scenario. For evaluating whether an adaptable operation is possible in adverse circumstances, e.g., by performing hybrid networking or graceful degradation, we finally use Machine Learning to generate a client-based QoS predictor and forecast the achievable QoS levels. 



\end{abstract}


\section{Introduction}
\label{sec:introduction}

Fast and reliable communication systems will be one of the major enablers for cooperative automated driving. Cellular vehicle-to-everything (C-V2X) communication, which is currently based on LTE and evolving towards 5G, has lately gained attention as a potential solution to this challenge, favored by the ubiquity of deployed cellular networks. Hybrid networking via the two operation modes currently supported by V2X further increases the reliability by using either the sidelink PC5 interface or ITS-G5 for short-range communications, while the uplink/downlink Uu interface via the Core Network (CN) will be used for an increased communication range \cite{RN11908}. This work focuses on the latter case. 

From the application perspective, cooperative self-driving systems have strict Quality of Service (QoS) requirements for data rate, reliability, and end-to-end (E2E) delay that need to be satisfied despite the highly-dynamic nature of vehicular networks, where instantaneous key performance indicators (KPIs) fluctuate rapidly. The degradation of such indicators leads to a worse QoS performance, restricting the use of safety-related applications. 
Previous works that evaluated the feasibility of using commercially-deployed LTE networks for such applications were based on a single LTE network \cite{RN11391,RN11935}, and did not consider performance differences among Mobile Network Operators (MNOs) in a city. 

Although identifying and selecting the best performing MNO increases the reliability of the communication, routing packets through the CN will unavoidably increase the delay overhead making it challenging for safety-related applications to fulfill their QoS requirements. A possible solution to this problem is to use Mobile Edge Computing (MEC) \cite{RN11924} to enhance the LTE architecture by enabling the access to cloud-based services directly from the Radio Access Network (RAN), without having to route any traffic through the CN. Under this paradigm, a reduction of the  average delay of up to $80\%$ has been reported using simulations \cite{RN11895}. However, no experimental work that confirms this has been published yet. 

An additional step for the realization of resilient autonomous systems is the introduction of real-time QoS prediction. This will allow applications to gracefully degrade their operation and adapt to future conditions, maintaining a necessary safety level at all times, while still providing the optimal performance based on the best achievable QoS. 

This QoS prediction can be done in different ways: Using a network-based, client-based, or combined approach. Network-based QoS predictors deployed in a MEC architecture provide on-demand notification messages informing about the foreseeable QoS based on current network KPIs. However, if the operator cannot support such functionality at all times, e.g., due to roaming, handovers, or being out of coverage, no prediction will be available. Further, only averaged statistics will be provided and fast changing situations cannot be taken into account. Client-based or UE-based QoS predictors, on the other hand, are intended to be deployed on the vehicle itself providing prediction based on the User Equipment's (UE) current KPI measurements and a trained Machine Learning (ML) model. Here, the vehicle will be capable of sensing the current network situation and predicting the achievable QoS level independently of the MNO.

Previous works on LTE QoS prediction have focused on forecasting throughput using either network-based \cite{cellular_prediction} or client-based data \cite{RN9825,RN963}. Existing literature on delay prediction neither takes into account dynamic channel conditions nor uses advanced methods such as  ML models \cite{volos2019reladec}. 

The main contributions of this paper are therefore threefold, addressing the aforementioned knowledge gaps:
\begin{enumerate}
  \item We expose and quantify differences among MNOs and assess their readiness to support C-V2X communications.
  \item We provide empirical results to evaluate the enhanced performance of a live MEC network deployed at the premises of one MNO, as well as its suitability for different V2X applications.
  \item We forecast E2E delay performance using a UE-based QoS predictor which is generated by training an ML model based on instantaneous KPIs. These KPIs were collected from live LTE networks on the road using a prototype that we implemented.
\end{enumerate}

\section{Measurement Setup and Data Collection}
\label{sec:measurement}

\subsection{Empirical Uu-based C-V2X Setup}

\begin{figure}
\centering
\begin{minipage}[t]{.49\linewidth}
  \includegraphics[width=\linewidth]{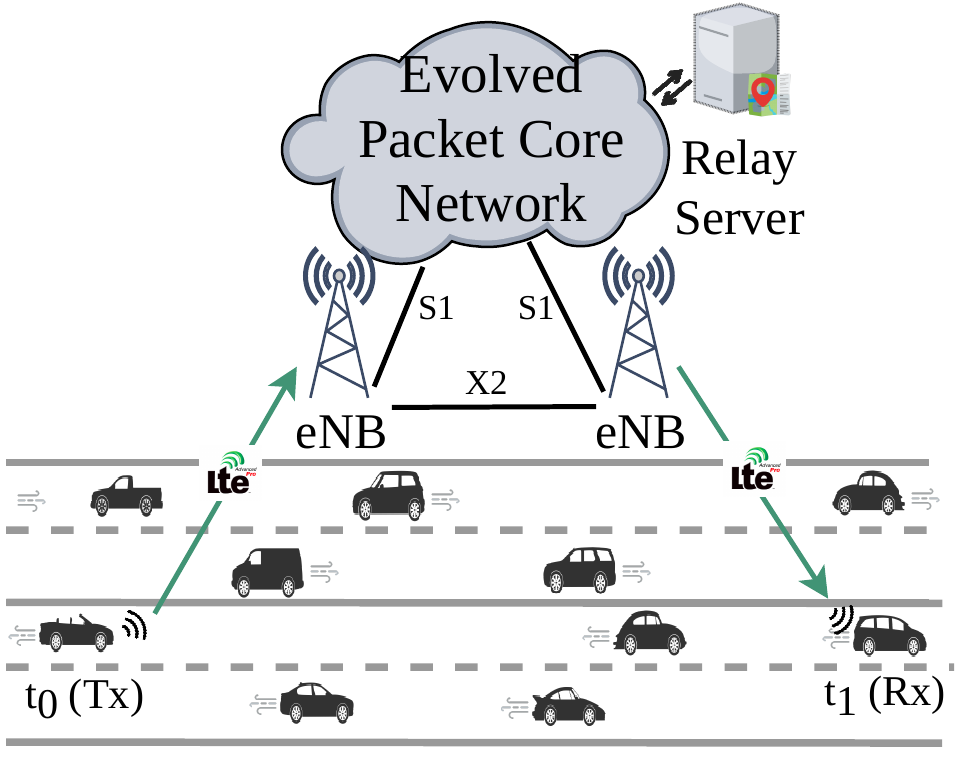} 
  \captionof{figure}{C-V2X scenario.}
  \label{fig:cv2x_scenario}
\end{minipage}
\hfill
\begin{minipage}[t]{.49\linewidth}
  \includegraphics[width=\linewidth]{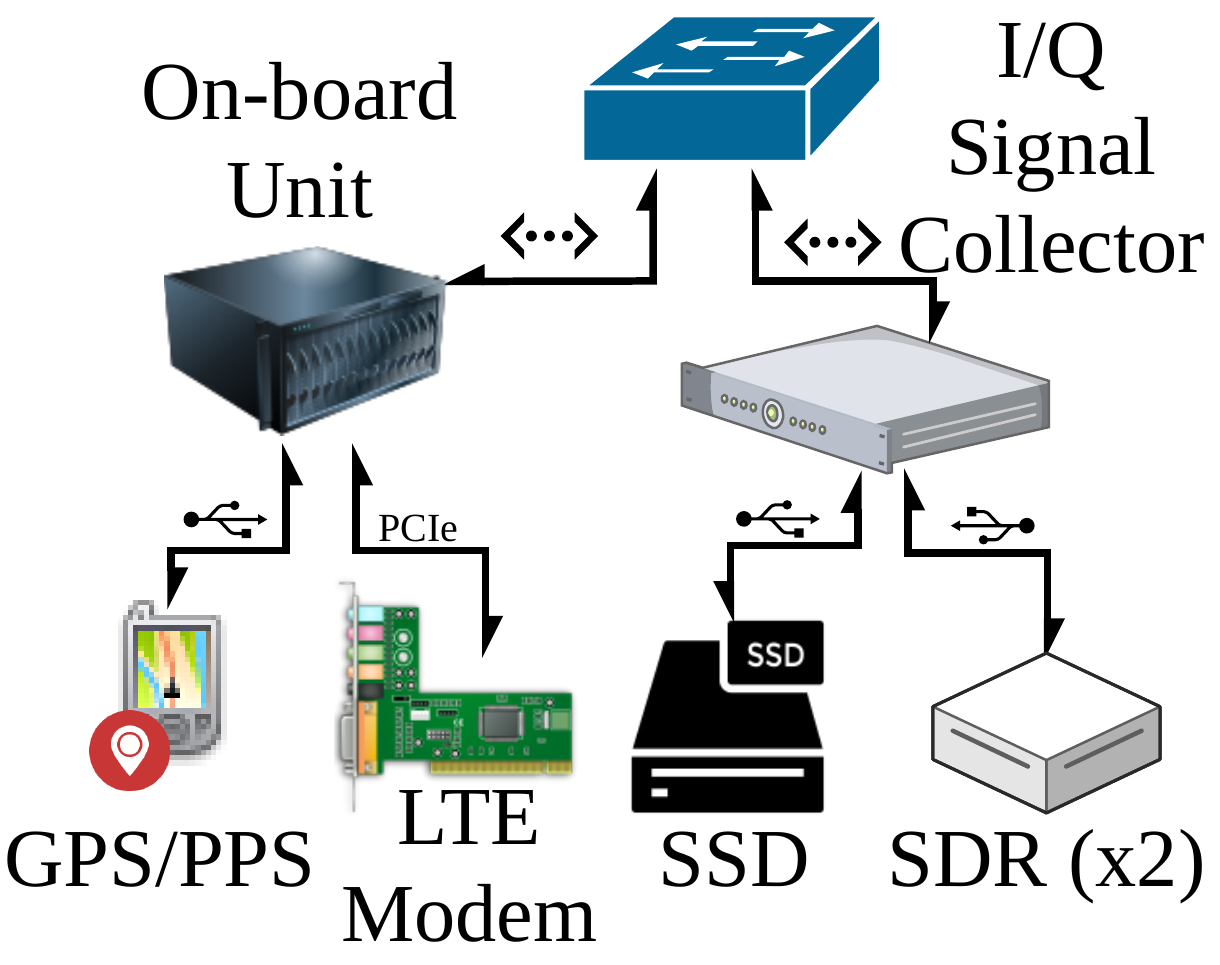} 
  \captionof{figure}{Hardware setup.}
  \label{fig:hw_setup}
\end{minipage}%
\end{figure}

The C-V2X scenario under consideration, illustrated in \figref{fig:cv2x_scenario}, shows how the communication link is established: The transmitting vehicle uses a commercially-deployed LTE network to send UDP datagrams to a relay server, which forwards the messages to the receiving vehicle connected to the same LTE network. Two relay servers have been deployed in our setup: One hosted on the Internet for assessing its access via commercially-deployed LTE networks, and the other located on MNO1's premises and thereby closer to its eNBs for evaluating a MEC network architecture.

Our measurement campaigns aimed to collect KPIs of multiple MNOs simultaneously in order to investigate a potential correlation between these metrics and the QoS in terms of the E2E delay, as well as their readiness for applications with strict QoS requirements. Here, we define the E2E delay as the time it takes to deliver a packet from the transmitting vehicle via the network and the relay server to a receiving vehicle. 
We covered an urban scenario with a mix of highways and inner-city streets, and transmitted messages with a fixed payload size and constant intervals. While different packet sizes were initially investigated, we finally chose a fixed packet size of $300$ bytes, since this is the average size of a ETSI Cooperative Awareness Message (CAM) \cite{ETSI_CAM}, a periodically broadcast message with vehicle's status information. As a transmission interval, we chose $40$ ms as further explained in \secref{sec:static}.


\subsection{E2E Delay Measurements using COTS equipment}

For setting up the C-V2X communication we developed a UDP traffic generator that runs on an On-Board Unit (OBU) equipped with the Commercial Off-The-Shelf (COTS) \textit{Sierra Wireless EM7565} LTE modem and connected to a roof-mounted antenna of our vehicle. For each of the three MNOs as well as the MEC architecture, a separate OBU was mounted in the vehicle, each with a SIM card restricted to access only LTE networks and configured with a network-specific Access Point Name (APN). Carrier aggregation was disabled at the UE. Since a single vehicle was used for the experiment, each OBU acted both as transmitter and receiver for a particular MNO, transmitting UDP datagrams with a unique sequence number as payload and registering the timestamps at which it was sent and received back from the relay server. Clock synchronization among OBUs was performed via the Precision Time Protocol (PTP) by a master OBU using the high-precision Pulse-Per-Second (PPS) signals provided by the \textit{u-blox EVK-M8} Global Navigation Satellite System (GNSS) receiver, which additionally collected geographical information along the road. \figref{fig:hw_setup} illustrates the equipment used per monitored MNO.

\begin{figure}
 \centering
 \includegraphics[width=.96\linewidth]{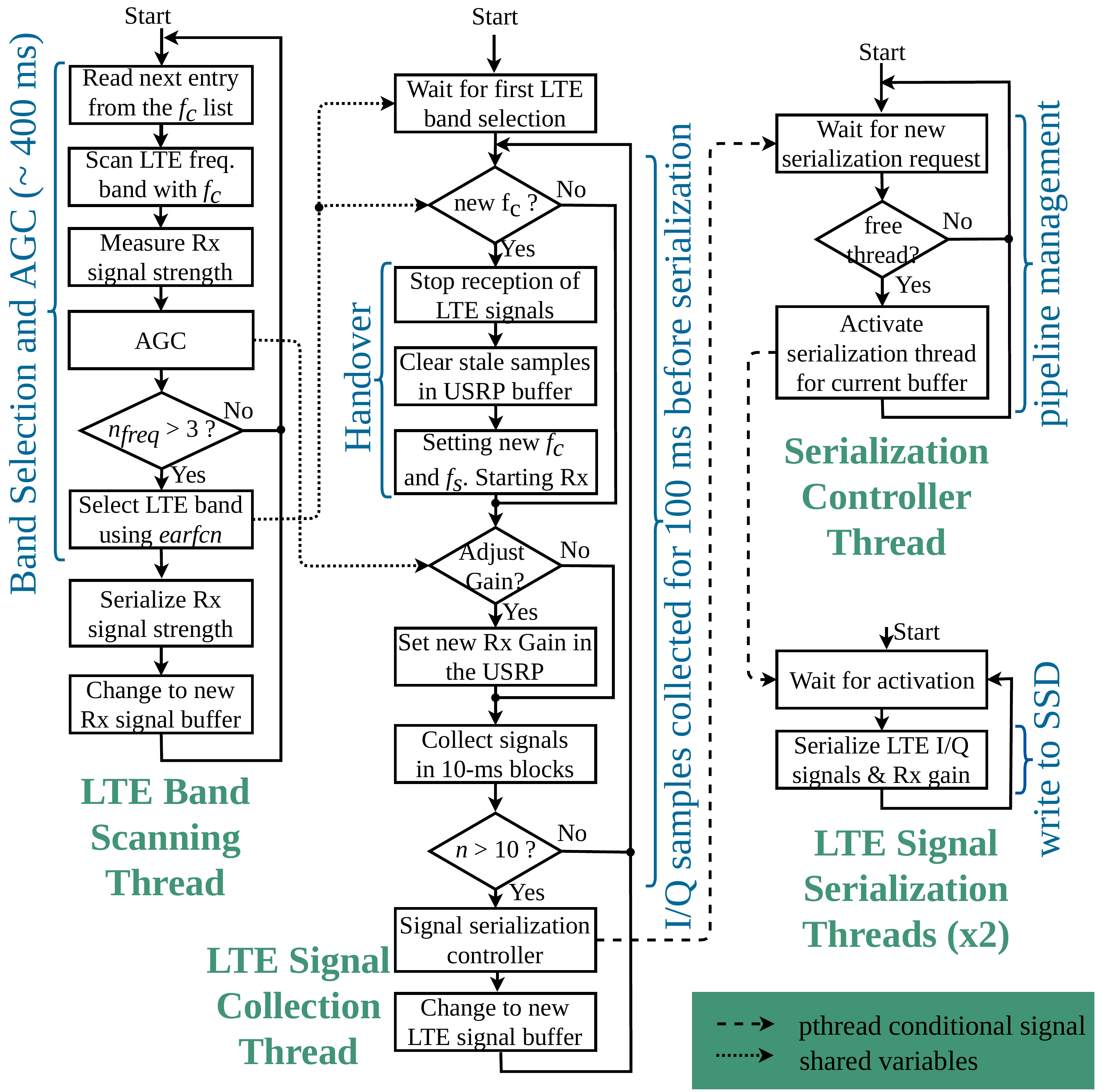}
 \captionof{figure}{Flowchart of the real-time \textit{I/Q Signal Collector}.}
 \label{fig:signal_collector}
\end{figure}

\subsection{Real-time I/Q Sample Collection}

One major issue of using COTS LTE modems for collecting KPIs was the introduction of hardware constraints that prevented precise KPI extraction: We detected identical readings over periods of $1$-$3$ seconds where signal fluctuations were evidently overlooked, making such readings unsuitable for precise QoS prediction. In order to increase the granularity of our measurements we implemented a \textit{Signal Collector}, whose design is depicted in \figref{fig:signal_collector} and main task is the collection of raw I/Q signals from the serving eNB for later post-processing. The \textit{Signal Collector} has been realized using five main threads. 

The \textit{LTE Band Scanning Thread} monitors downlink channels and estimates the receive gain needed in each channel for successful LTE signal decoding. For this, it uses an Automatic Gain Control (AGC) tuned for our \textit{USRP B210} Software-Defined Radio (SDR). Additionally, it identifies the serving eNB's downlink channel based on the E-UTRA Absolute Radio Frequency Channel Number (\textit{earfcn}) polled by the OBU from its LTE modem at all times and made available via LAN. The SDR controlled by this thread operates in discontinous mode to avoid buffer flushing, thus speeding up band scanning.

The \textit{LTE Signal Collection Thread} uses the previously discussed information to tune to the channel used by the COTS modem and start collecting LTE signals in blocks of 10 ms with the appropriate carrier frequency ($f_c$), sampling rate ($f_s$) and receive gain. Further, it monitors the eNB's $f_c$ every \mbox{10 ms} to determine if a handover has occurred. To ensure a real-time operation, an SDR running in continuous mode is used here. After collecting ten blocks of 10 ms, an alternative buffer is allocated for measurements over the next 100 ms, allowing the \textit{Serialization Controller Thread} to use the inactive buffer for writing its content to a solid-state disk (SSD), which is ultimately carried out by one of the two available \textit{LTE Signal Serialization Threads} in order to parallelize such operation.


The integration of such system with the OBU is done using an on-board switch as shown in \figref{fig:hw_setup}, while co-located roof antennas connected to the SDRs and LTE modem ensured that the received signals underwent similar channel effects.




\subsection{Decoding raw LTE Signals to Extract KPIs}

After concluding the measurements, the collected raw I/Q samples are post-processed using a modified version of the UE implementation of OpenAirInterface (OAI-UE) \cite{oai}, which is used to perform downlink LTE PHY-layer procedures \cite{3gpp.36.213} on the de-serialized I/Q signals, and thus obtaining specific KPIs, as defined in \cite{3gpp.36.214}. In order to ensure the reliability of the post-processed information, false positives are removed based on an intermediate verification step consisting in doing a one-to-one mapping between the Physical-layer Cell Identifier (PCI) decoded using our offline version of OAI-UE and the E-UTRAN Cell Identifier (ECI) provided by the LTE modem. For this purpose, a maximum likelihood estimation approach is used for the time window when the UE was connected to each cell. The information obtained from the different data sources is finally merged into a unique overall data set based on the aforementioned GNSS-based clock synchronization among on-board systems. The list of relevant KPIs collected along the road is shown grouped by source in \tabref{tab:kpi_measurements}.





\begin{table}[t]
 \begin{center}
\begin{tabular}{ c c c }
\toprule
 Source & KPIs & Granularity \\ 
 \midrule
 \begin{tabular}{@{}c@{}} COTS LTE \\ modem \end{tabular} & \begin{tabular}{@{}c@{}} \textit{SINR, RSSI, RSRP, RSRQ,} \\\textit{LTE band, earfcn, ECI}\end{tabular} & $\sim1$-$3$ s  \\
 \midrule
 GPS receiver & \textit{latitude, longitude, speed} & $50$ ms \\
 \midrule
 OAI-UE & \begin{tabular}{@{}c@{}} \textit{SINR, RSSI, RSRP, RSRQ,} \\\textit{Noise Power, Rx Power, PCI} \end{tabular} & $10$ ms \\ 
 \bottomrule
\end{tabular}
\caption{KPIs collected during the drive test.} \label{tab:kpi_measurements}
\end{center}
\end{table}

\begin{figure}
 \centering
 \input{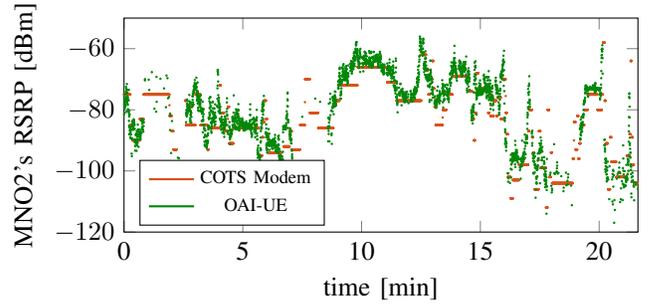}
 \captionof{figure}{OAI vs. COTS LTE modem measurements.}
 \label{fig:w27_oai_vs_modem}
\end{figure}

In order to evaluate the validity of our OAI-UE approach, we compare the Reference Signal Received Power (RSRP) values obtained by decoding raw signals with the measurements obtained using the COTS LTE modem. \figref{fig:w27_oai_vs_modem} shows this comparison for an exemplary snapshot of a few minutes for MNO2, which indicates that both approaches agree on the overall behavior, except for a few outliers. Further, the figure clearly shows the different resolution in time of the COTS LTE modem and the OAI-UE. Due to the higher update rate, the OAI-UE measurements are more beneficial for a packet level QoS prediction; therefore, we will focus on OAI-UE KPIs in the remainder of the paper.


\subsection{Effect of the Transmission Interval in the E2E Delay}
\label{sec:static} 

\begin{figure}
  \centering
  \input{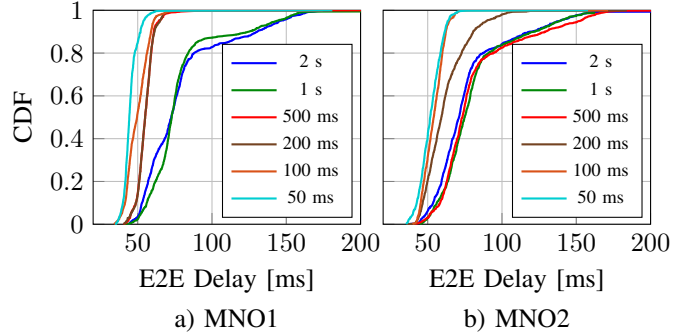}
  \caption{Effect of different transmission intervals on the E2E delay in a static setup, for a payload size of $100$ bytes.}
  \label{fig:inlab_tx_interval_test}
	\vspace{0.5\baselineskip}
\end{figure}

Based on preliminary measurements, we show that different transmission intervals have a significant impact on the E2E delay, as depicted in \figref{fig:inlab_tx_interval_test}. These results show that messages transmitted at shorter intervals, e.g., every $50$ ms, reduce the average E2E delay by $34.34$ ms (MNO1) and $27.68$ ms (MNO2)
compared to those transmitted at longer intervals, e.g., every $1$ s. This suggests that for any outgoing packet the resulting E2E delay is influenced by the time elapsed since the last transmission to the same destination. This is likely caused by the various routing protocols at the core and ISP networks or vendor-specific features of the eNBs (e.g., the scheduler's algorithm) which are out of the scope of this work. 
Nevertheless, in order to prevent these factors from strongly influencing the E2E delay we use a fixed transmission interval lower than this threshold, i.e., $40$ ms, given that these effects have a minor impact below $50$ ms.

\begin{figure} [t]
 \centering
 \input{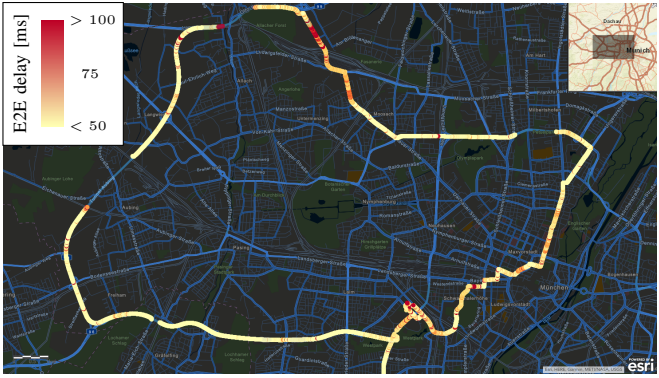}
 \caption{Locations of MNO3's delay measurements, indicating the route covered in Munich's urban area.} 
 \label{fig:user_density_scenarios}
\end{figure}

\section{Data Analysis and Results} 
\label{sec:data_analysis}


\subsection{Adequate Allocation of Radio Resources} 
\label{sec:dt2_data_analysis_kpi}


Two main eNB design criteria are used in urban scenarios, such as the one studied herein and shown in \figref{fig:user_density_scenarios}: \mbox{(1) high user density} favors the deployment of closely-located eNBs where the use of high frequencies is preferred due to their stronger channel attenuation, thus reducing the coverage area of a cell and increasing the Signal-to-Interference-plus-Noise Ratio (SINR) in the adjacent cells; and (2) higher traffic demand justifies the use of wider bandwidth increasing the cell capacity and the number of users that can be simultaneously served.

We found that MNO1 mostly fulfilled these two criteria, while we observed deviations for MNO2 and MNO3. In the case of MNO2, our vehicle spent $24\%$ of the time in coverage of $25$ eNBs configured with a carrier frequency ($f_c^\text{DL}$) more suitable for rural areas, while $65.7\%$ of the time it was served by $10$ MHz instead of $20$ MHz bandwidth cells. Slightly better results were obtained for MNO3. 
Thus, these eNBs in MNO2 and MNO3 will experience worse performance for C-V2X communications under crowded conditions, unless radio resource optimizations are conducted.

The overall frequency mapping results are shown in \tabref{tab:earfcn_list}, which also lists the LTE channel configurations detected for each MNO including the percentage of the transactions collected in each channel. The total number of eNBs covered in our drive test were $83$, $114$, and $86$ for MNO1, MNO2, and MNO3, respectively.

\begin{table}[t]
	\centering
	\begin{tabular}{c c c c c c}
		\toprule
		MNO & Band & \textit{earfcn} & $f_c^\text{DL}$ & Bandwidth & Percent.\\
		\midrule
		\multirow{3}{*}{MNO1} & 8 & 3749 & 954.9 & 10 MHz & 6.1\% \\
		 & 3 & 1300 & 1815 & 20 MHz & 79.6\% \\
		 & 3 & 1444 & 1829.4 & 10 MHz & 14.3\% \\
		\midrule
		\multirow{4}{*}{MNO2} & 20 & 6300 & 806 & 10 MHz & 24\% \\
		 & 3 & 1801 & 1865.1 & 20 MHz & 34.3\% \\
		 & 1 & 101 & 2120.1 & 10 MHz & 34.4\% \\
		 & 7 & 2850 & 2630 & 10 MHz & 7.3\% \\
		\midrule
		\multirow{3}{*}{MNO3} & 3 & 1600 & 1845 & 10 MHz & 55.2\% \\
		 & 1 & 252 & 2135.2 & 20 MHz & 37.9\% \\
		 & 7 & 3350 & 2680 & 20 MHz & 6.9\% \\
		\bottomrule
	\end{tabular}
	\caption{LTE channels detected during drive test, including percentage of occurrence. Carrier frequency ($f_c^\text{DL}$) in MHz. Sample size: MNO1: $\sim$105k; MNO2: $\sim$133k; MNO3: $\sim$100k.}
	\label{tab:earfcn_list}
\end{table}



\subsection{Radio Conditions along the Road} 

The allocation of sufficient radio resources does not necessarily guarantee a good QoS performance in C-V2X communications. Adequate transmission power management and handover mechanisms at the eNBs are also needed for optimal operation of the RAN. Consequently, these two factors have a large impact on the KPI measurements at the UE, which are later fed back to the eNB for selecting the Modulation and Coding Scheme (MCS). The MCS in turn determines the achievable throughput and E2E delay of the cell. Thus, having poor signal reception either by lack of coverage or by inefficient handover mechanisms is linked to poor QoS.

\begin{figure}
  \centering
  \input{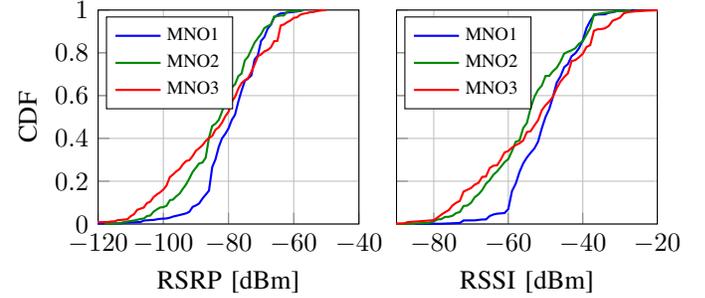}
  \caption{Measurements of RSSI and RSRP KPIs.}
  \label{fig:w27_kpi_measurements}
\end{figure}

\figref{fig:w27_kpi_measurements} shows a comparison of the MNOs based on the probability distribution of their RSRP and Received Signal Strength Indicator (RSSI) as measured during our drive test. This figure clearly shows that MNO3 has a significantly higher portion of measurements that were taken in cells with poor coverage, followed by MNO2. In contrast, MNO1 exhibits a much better KPI distribution, where just $8.7\%$ of the measurements present fair-to-poor coverage \mbox{($\mathrm{RSRP} \leq -90$ dBm)}, compared to $26.3\%$ for MNO2 and $34.2\%$ for MNO3. These differences will influence the QoS performance for each operator.

\subsection{Overall E2E Delay Performance for V2X Applications}
\label{sec:dt2_data_analysis_delay}

\begin{figure}
 \centering
 \input{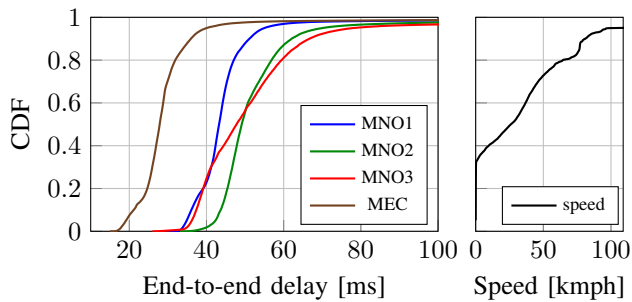}
 \captionof{figure}{Speed and E2E delay distributions for all MNOs.}
 \label{fig:w27_delay_distributions}
\end{figure}

Our measurements, shown in \figref{fig:w27_delay_distributions}, confirm that the previously discussed stronger signal reception at the UE exhibited in MNO1's RAN network is linked to a lower E2E delay, in contrast to MNO2 or MNO3 where lower RSRP/RSSI levels correlate to a worse performance. Here, stronger signal strength measurements encourage choosing higher-order modulation schemes as well as higher coding rates, which improve the throughput at the expense of weakening the error protection. As a result, the transmission delay is shortened which in turn decreases the overall E2E delay.

In this paper, we evaluate two main V2X applications: Advanced driving (safety-related application) and Day-1 use cases (e.g., Emergency Brake Light \cite{eu_cits_rep}), with an E2E delay requirement of $50$~ms and $100$~ms, respectively \cite{RN11912,RN5782,delay_50ms_req,delay_100ms_req}. These QoS requirements are used to evaluate whether the LTE networks under study are able to support such applications. The compliance for both use cases is presented in \tabref{tab:mno_v2x_app_qos_compliance}, where the portion of packets that fulfills each delay requirement is shown for each MNO and the MEC case.

These results, which are also presented in \figref{fig:w27_delay_distributions}, show that in spite of MNO1's superior performance compared to the other two operators, it does not comply with the requirements of the safety-related application in $12.2\%$ of the cases. This, however, improves as we analyze Day-1 use cases, where the E2E delay requirement rises to $100$ ms, and a performance above $96\%$ is achieved for all MNOs, making all studied LTE networks potentially suitable for Day-1 applications. 

\subsection{Evaluation of the E2E delay using a MEC Architecture}

The deployment of the MEC architecture in the network of MNO1 aims to further reduce the E2E delay. As a result, the delay shown in \figref{fig:w27_delay_distributions} is reduced by $18.46$ ms on average with respect to the non-MEC case. Further, MEC achieves $97.6\%$ and $98.7\%$ of the cases a delay below $50$ and $100$ ms, respectively, as shown in \tabref{tab:mno_v2x_app_qos_compliance}. 

\begin{figure}
	\centering
	\input{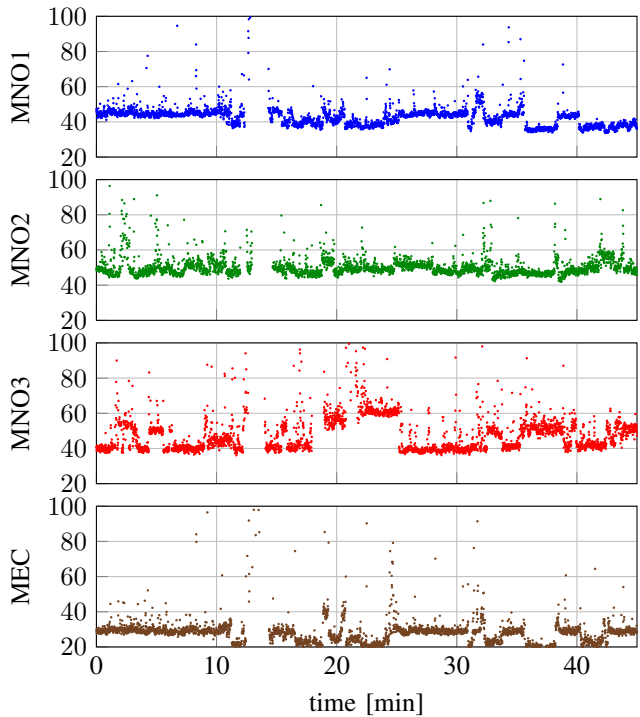}
	\captionof{figure}{E2E delays over a time snapshot for all MNOs/MEC.}
	\label{fig:w27_delay_time}
\end{figure}

Despite the improved performance of the MEC case, $2.4\%$ of the packets do not adhere to the $50$ ms threshold. This can be seen more clearly in \figref{fig:w27_delay_time}, which shows the E2E delay in the time domain for all cases. For visual clarity, samples have been averaged over time windows of $250$ ms. The plot shows short bursts of delays above the $50$ ms threshold at different times and for short durations, which likely points to inefficient handovers or different parameter configurations in the eNBs. Further, this indicates that while one eNB may be very reliable for safety-critical C-V2X communication, the next one may be very unreliable. Finally, there are out-of-coverage periods such as when traversing a tunnel, e.g., around the $13$ minute mark. These effects are also noticeable in \figref{fig:user_density_scenarios} which shows the delays of MNO3 along the covered route: Darker zones (red) represent underperforming eNBs or non optimal handover points where the delay increases, while lighter zones (yellow) have the potential of supporting Day-1 V2X applications. Gaps in the route relate to traversed tunnels.

Our vehicle's speed distribution is shown in \figref{fig:w27_delay_distributions} for the sake of completeness. The average speed was $28.4$ kmph.

\begin{table}[t]
 \begin{center}
\begin{tabular}{c c c c c c}
\toprule
 Use Case & Delay & MNO1 & MNO2 & MNO3 & MEC\\ 
 \midrule
 \begin{tabular}{@{}c@{}} Day-1 \\ use cases\end{tabular} & $\leq100$ & $98.5\%$ & $97.6\%$ & $96.7\%$ & $98.7\%$ \\
 \midrule
 \begin{tabular}{@{}c@{}c@{}} Advanced\\driving \end{tabular} & $\leq50$ & $87.8\%$ & $55.3\%$ & $56.2\%$ & $97.6\%$ \\
 \bottomrule
\end{tabular}
\caption{Compliance of observed LTE networks with the QoS requirements for V2X applications. E2E delay given in ms.}
\label{tab:mno_v2x_app_qos_compliance}
\end{center}
\end{table}

\begin{figure}
	\centering
	\begin{tikzpicture}

\begin{groupplot}[
group style={
	group name=cdf_plots,
	group size=1 by 4,
	vertical sep=0.3cm,
	ylabels at=edge left,
	xlabels at=edge bottom,
	yticklabels at=edge left,
	horizontal sep=0pt,
},
width=0.99*\columnwidth,
height=0.39*\columnwidth,
axis on top,
xtick=\empty,
xmin=1,
xmax=10,
ymajorgrids,
ymin=20,
ymax=100,
extra y tick style={grid=major},
ylabel={E2E delay},
tickpos=left,
ytick align=inside,
xtick align=inside,
]

\nextgroupplot [ylabel={MNO1}]

\addplot graphics [xmin=1, xmax=10, ymin=19,ymax=101] {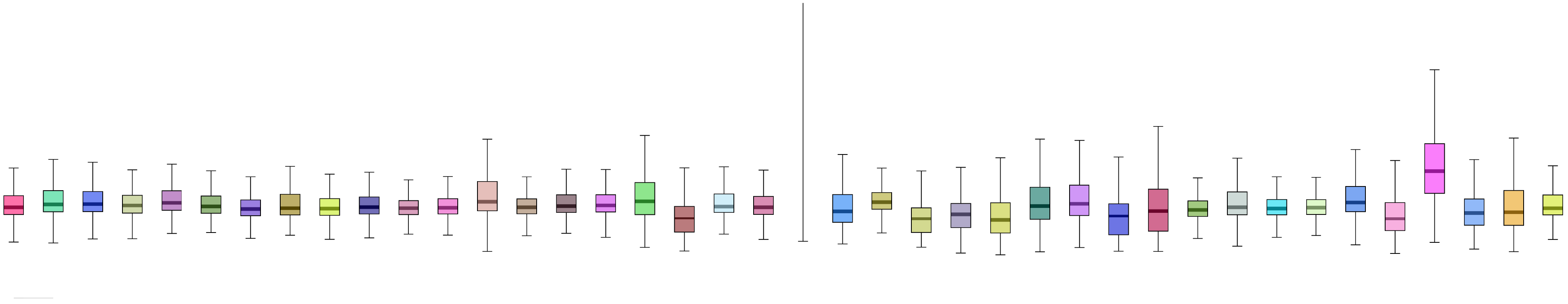};

\nextgroupplot [ylabel={MNO2}]

\addplot graphics [xmin=1, xmax=10, ymin=19,ymax=101] {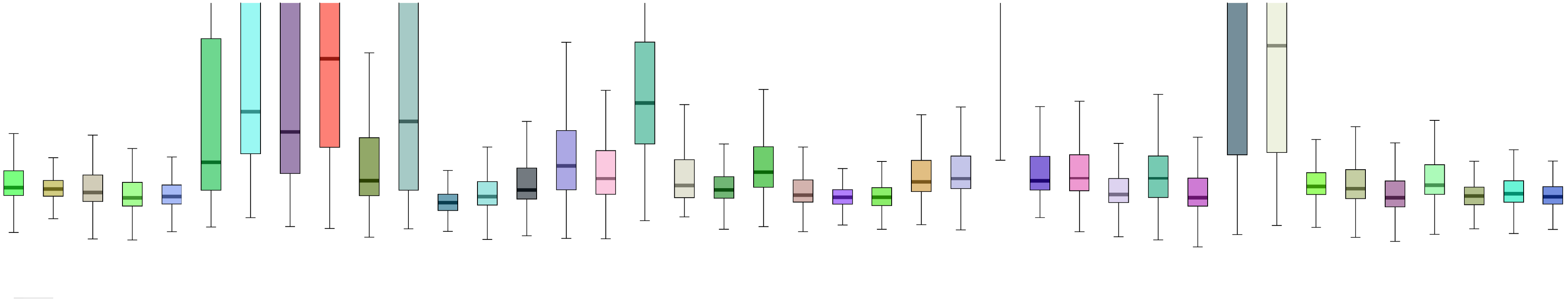};

\nextgroupplot [ylabel={MNO3}]

\addplot graphics [xmin=1, xmax=10, ymin=19,ymax=101] {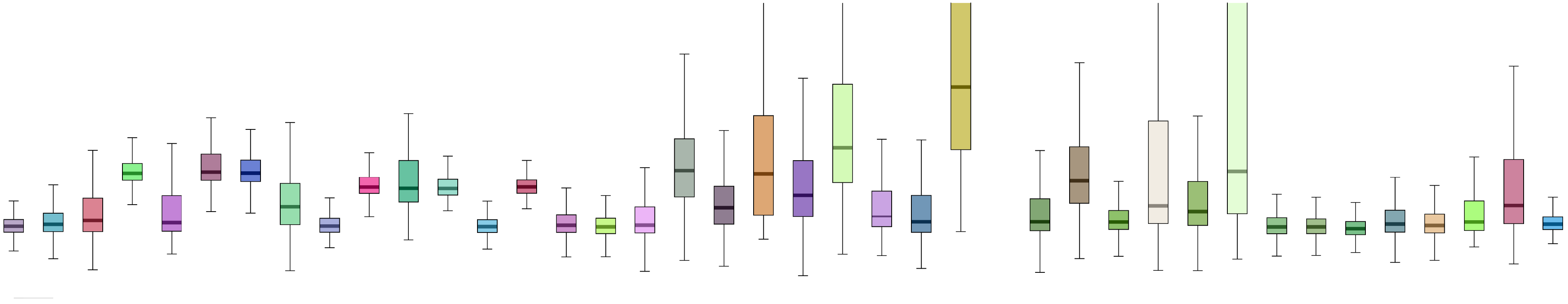};

\nextgroupplot [ylabel={MEC}]

\addplot graphics [xmin=1, xmax=10, ymin=19,ymax=101] {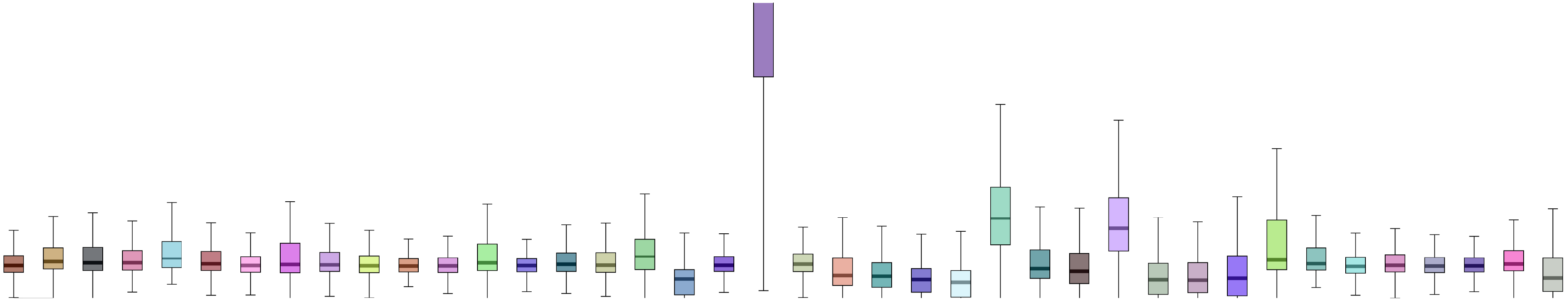};

\end{groupplot}


\end{tikzpicture}
	\captionof{figure}{Delay distribution per eNB sector for MNOs/MEC.}
	\label{fig:w27_eci_bloxplot_mno1}
\end{figure}

The underperforming eNB sectors can be more easily identified in \figref{fig:w27_eci_bloxplot_mno1}, which shows the E2E delay distribution per eNB sector for the first $40$ traversed eNB sectors for each MNOs and the MEC case. This shows that the dependence of the QoS performance on the current eNB has to be considered for safety-critical communications, as not all eNBs will meet the requirements with a high probability. Further, it highlights the need for a good prediction, the use of hybrid networking, or ultimately graceful degradation of the application.

While our analysis focuses on applications exchanging CAM messages, our results imply that Descentralized Environmental Notification Messages (DENM) \cite{ETSI_DENM} carrying event-driving hazard warning are strongly supported by existing LTE networks given their downlink-only nature.

\section{E2E Delay Prediction}
\label{sec:prediction}

Based on the previously discussed measurements, we developed a QoS prediction for the E2E delay of a new packet transmission. From the application perspective, the precise value is likely not as important as whether or not the delay is sufficiently low, i.e., below a given threshold. Therefore, herein we view the prediction as a classification problem of the E2E delay regarding a given threshold. In practice, the appropriate threshold depends on the target application, and thus would have to be tuned for each application class separately. In this paper, we consider the thresholds of $50$ and $100$ ms.

For the predictor we focus on supervised machine learning in the form of a Neural Network (NN) for classification, as NNs are able to learn very complex functions based on the provided features. 
The structure of the NN that we use is a single hidden layer with $128$ neurons, with the parameters determined using hyperparameter tuning for our data set. Further, in order to verify the training performance across the entire data set, we apply $k$-fold stratified cross-validation using $k=5$ \cite{k_fold}.


For the data set, we combine the non-MEC measurements of MNO1 and MNO2 to generate a large data set with a good delay spread and to avoid learning MNO-specific behavior. While it would be possible to differentiate between the MNOs based on the name of each eNB, we will neither consider the eNB name nor the MNO as a feature for learning and treat both MNOs as belonging to the same scenario. Further, we choose to omit position information for training in order to generate a prediction model which can extrapolate to unknown scenarios.

As previously mentioned, the E2E delay is not solely based on the RAN, but the CN and ISP networks as well. However, as the UE KPIs only give an indication of the performance on the RAN, we do not have any direct indication of delay occurring on CN and ISP networks. In order to mitigate this effect, we add the \textit{expected E2E delay} as a feature, i.e., the average delay of all packets transmitted per eNB during our measurements. While this feature is not readily available from COTS modems, it can either be crowdsourced or computed locally based on delay measurements for the serving eNB, which the application will likely monitor anyway. In the latter case, the \textit{expected E2E delay} feature will take a few packets to converge on an accurate value, which should only be a problem in the case of a very frequent handovers. 

In order to determine a good set of features for training the NN, we apply the Maximum Dependency (MD) algorithm \cite{MD} for feature selection. This algorithm iteratively builds the set of features, such that it maximizes the multivariate mutual information of the classifier, while avoiding redundant information. For this joint data set, the best set of features according to the MD algorithm in order of importance are \textit{expected E2E delay}, \textit{speed}, \textit{SINR}, \textit{RSRP} and \textit{RSSI}.

\begin{table}[t!]
 \centering
 \begin{tabular}{c | c  c  c  c}
  & precision & recall & f1-score & support \\
  \hline
  $\le 50$~ms & $0.8510$ & $0.9203$ & $0.8843$ & $0.7773$ \\
  \hline
  $> 50$~ms & $0.6109$ & $0.4372$ & $0.5090$ & $0.2227$ \\
 \end{tabular}
 \caption{Prediction quality using the joint data set.}
 \label{tab:binary_class_joint}
 \vspace{0.7\baselineskip}
\end{table}
\begin{table}[t!]
 \centering
 \begin{tabular}{c | c  c  c  c}
  & precision & recall & f1-score & support \\
  \hline
  $\le 50$~ms & $0.7611$ & $0.6954$ & $0.7267$ & $0.5$ \\
  \hline
  $> 50$~ms & $0.7196$ & $0.7817$ & $0.7494$ & $0.5$ \\
 \end{tabular}
 \caption{Prediction quality after balancing the classes.}
 \label{tab:binary_class_joint_balanced}
 \vspace{0.7\baselineskip}
\end{table}

In \tabref{tab:binary_class_joint} we show results of the prediction using binary classification for a threshold of $50$ ms. Here, we see that the delays $\le 50$~ms have a much higher f1-score, which is mainly due to the larger support. Further, while the overall accuracy is $\approx 0.8127$, the main problem is the precision of delays $\le 50$~ms, which means that approximately $15\%$ of all positive predictions for this threshold are incorrect. If we try to reduce the influence of different supports in both classes, by randomly removing a sufficient number of samples from the $\le 50$~ms class, such that both classes have the same support, the results change to those shown in \tabref{tab:binary_class_joint_balanced}. Due to this change, both classes have an f1-score of around $0.73$, while delays $\le 50$~ms have a higher precision and delays $> 50$~ms have a better recall, and the overall accuracy is $\approx 0.7385$. With this approach the misclassifications are more balanced, but the overall performance is not improved.

\begin{table}[t!]
 \centering
 \begin{tabular}{c | c  c  c  c}
  & NN & RNN & RF & SVM \\
  \hline
  joint data set & $0.8127$ & $0.8176$ & $0.7967$ & $0.8107$ \\
  \hline
  balanced data set & $0.7385$ & $0.7387$ & $0.7018$ & $0.7372$ \\
 \end{tabular}
 \caption{Prediction accuracy using different ML approaches.}
 \label{tab:ml_comp}
 \vspace{0.7\baselineskip}
\end{table}

In order to verify if the chosen type of ML algorithm is a good choice, we compare the prediction accuracy of different common ML approaches for the joint data set, as well as the balanced version of this data set in \tabref{tab:ml_comp}. In addition to the NN, we consider a Recurrent Neural Network (RNN) with Long Short-Term Memory (LSTM) neurons, a Random Forest (RF), and a Support Vector Machine (SVM). From the results in \tabref{tab:ml_comp} it is clear that none of the investigated approaches performs significantly better than the NN, and only the RNN has a marginally better prediction accuracy. This is expected since the RNN is well suited for time sequences and high E2E delays often occur in bursts. However, this slight increase in accuracy does not justify the increased cost of training and execution of the RNN compared to the much simpler NN.

\begin{table}[t!]
 \centering
 \begin{tabular}{p{1.7cm} | c  c  c  c}
  & precision & recall & f1-score & support \\
  \hline
  $\le 50$~ms & $0.7991$ & $0.9324$ & $0.8606$ & $0.7056$ \\
  \hline
  \parbox[c]{1.7cm}{\strut $> 50$~ms \& $\le 100$~ms \strut} & $0.4519$ & $0.3898$ & $0.4181$ & $0.1944$ \\
  \hline
  $> 100$~ms & $0.6579$ & $0.05807$ & $0.1053$ & $0.1000$ \\
 \end{tabular}
 \caption{Prediction quality for multiclass classification.}
 \label{tab:multiclass}
\end{table}
As multiple V2X applications may share the same wireless link, it is important to also consider the prediction of multiple thresholds. To this end we consider the prediction performance of three classes: \textquote{$\le 50$~ms}, \textquote{$> 50$~ms \& $\le 100$~ms}, and \textquote{$> 100$~ms}. In our measurements delays of $> 100$~ms only occurred in $\approx 2 \%$ of all packets, which is too low for reliable training. Therefore, to generate a more balanced data set we randomly removed packets from the other two classes, such that these two classes keep their ratio the same and the $> 100$~ms class has relative support of $10\%$. \tabref{tab:multiclass} shows prediction performance for this multiclass classification approach. With an accuracy of $\approx 0.7395$, these results show similar issues to the binary classification approach: Those classes which do not occur often have a low f1-score and the overall accuracy is not very high.



\section{Conclusion}
\label{sec:conclusion}

In this paper, we evaluated the performance of C-V2X communications by carrying out on-road measurements for multiple MNOs. Our results show that existing LTE networks are suitable to support Day-1 applications with delay requirements below $100$ ms up to $98.5\%$ of the time, while more critical applications demanding delays below $50$ ms will still face difficulties. 
Introducing a MEC architecture addressed the latter requirements in $97.6\%$ of the cases, reducing the average E2E delay by $18.46$ ms. Further, we identified underperforming eNBs and handovers that will potentially account for a higher performance if optimized. Thus, current achievable delay is partially limited by LTE's radio access network; 5G has a higher potential for more demanding future use cases. 



In order to adapt our V2X application to varying QoS levels, we also investigated how well the collected data is suited for achieving a reliable QoS prediction. With our prediction approach we achieved f1-scores of up to 88\% without using any position information or MNO-specific optimizations. This is unfortunately not sufficient for highly reliable safety-critical applications. While adding further measurements would improve the prediction performance, it is uncertain if this results in a highly accurate prediction. The main issue is that the measured UE KPIs did not give a sufficiently clear indication of the resulting E2E delays in all situations. A major reason being that no information regarding the CN and ISP networks or the eNB configuration was available. This indicates that a combined network- and UE-based QoS prediction approach which incorporates information from the MNO, e.g., new features provided by a network-based QoS predictor, should lead to a significantly more reliable prediction.



Our results show a great challenge for improving current LTE and future 5G networks, and the need for reliable QoS prediction to enable hybrid networking or graceful degradation of the application. 




\section*{Acknowledgement}

This work has been in part funded by the Bavarian Ministry of Economic Affairs as part of the \textit{Connected Mobility - Intelligent Vehicle Networking} project.



\bibliography{qos_prediction}
\bibliographystyle{ieeetr}

\end{document}